\documentclass[aps,prl,reprint,twocolumn,groupedaddress]{revtex4}
\usepackage{graphicx}% Include figure files
\usepackage{bm}        %added by L.Zhou
\usepackage{amsmath}   %added by L.Zhou 03/09/10
\begin{document}

%\preprint{}

\title{Lattice relaxation of dimer islands on Ge(001) during homoepitaxy by pulsed laser deposition}


\author{Lan Zhou}
\author{Yiping Wang}
\altaffiliation[Present address: ]{Nanjing University of Aeronautics and Astronautics, Nanjing, China.}
\author{Minghao Li}
\author{Randall L. Headrick}
\email[]{rheadrick@uvm.edu}
%\thanks{}
\affiliation{Department of Physics and Materials Science Program, University of Vermont, Burlington, Vermont 05405}


%\date{\today}
%\setcounter{page}{76}

\begin{abstract}
In low-temperature pulsed growth two-dimensional islands form and coarsen into $\sim$10 nm features. The islands produce well-defined displaced x-ray diffraction peaks due to relaxation of anisotropic surface stress of the (2$\times$1) reconstruction with expansion and contraction present in orthogonal directions.   We infer that the island distribution differs from continuous deposition, enhancing the population of size selected islands exhibiting anisotropic relaxation.
  The relaxation carries over into multilevel islands, suggesting that  domains in subsequent layers form metastable stress domains. \end{abstract}

%\keywords{}

\maketitle



Surface stress plays a   role in determining the equilibrium structure of clean surfaces, as well as the evolution of structures during epitaxial growth.\cite{Ibach1997} Many surfaces  exhibit anisotropic surface stress, which can influence the shapes of  two dimensional islands, mound structures, and mesoscopic facets.\cite{Alerhand1988, Kochanski1990, Li2000,  Middel2002} Even on surfaces where the lowest   energy is a flat surface, crystal growth at low temperature can reveal anisotropic effects through the appearance of low symmetry nonequilibrium features. Improved understanding of stress effects on surface growth may  lead to approaches for use of elasticity as a tool for self-organization. 
%In addition, such effects may play a role in phenomena such as limited thickness epitaxial breakdown observed during low temperature homoepitaxy.\cite{Eaglesham1990, Xue1993, Bratland2003}

\begin{figure}[htbp]
   \centering
   \includegraphics[width=2.8 in]{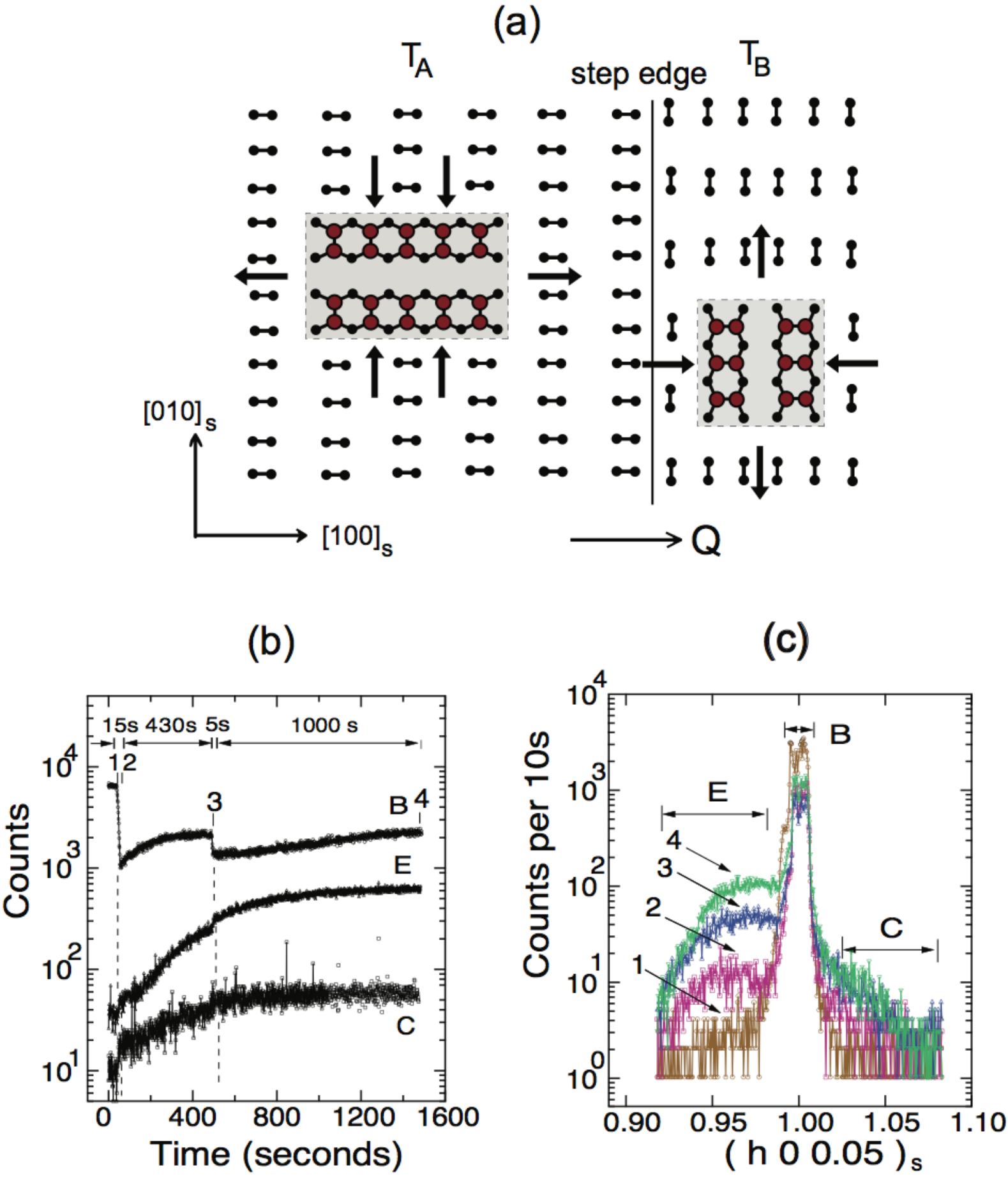}
   \caption{\footnotesize GID evolution of (1 0 0.05)$_s$ reflection during sub-monolayer growth on Ge(001) by PLD at 100$^\circ$C. (a) Schematic top view of small 2D dimer islands nucleated on a reconstructed surface containing degenerate (2$\times$1) and (1$\times$2) domains. Solid line represents the mono-atomic height step and big (red) circles are the dimer atoms. The arrows represent the elastic forces along the island periphery arising from the intrinsic surface stress anisotropy. The x-ray diffraction vector $\vec{Q}$ is oriented in the direction across the steps resulting from surface miscut. (b) The evolution of peaks E, B and C as a function of time.  Dashed lines represent laser bursts: 1st,  150 pulses (0.3 ML), followed by 430 s recovery; 2nd,  50 pulses to a total coverage of 0.4 ML followed by 1000 s recovery. The laser repetition rate is 10 Hz and the deposition rate is 0.002 ML/pulse. (c) Diffraction profiles: (1) starting surface; (2) right after 1st laser burst; (3) before the 2nd laser burst; (4) after the 2nd laser burst and recovery.  Note that the E peak is consistent with islands nucleated on $T_A$ terraces while C corresponds to islands atop $T_B$ terraces. }
   \label{fig:sub_ML}
\end{figure}

On Ge(001) surfaces, neighboring surface atoms dimerize to form a (2$\times$1) reconstruction to minimize the surface energy, but induce a large anisotropic stress.\cite{Zandvliet2003} The stress is tensile parallel to the dimer bond, and compressive normal to it. Experimental support for stress anisotropy from the shape of islands, vacancy clusters and domain wall fluctuations have been reported.\cite{Wu1995, Zandliet1999, Li2000,  Middel2002} In this work, we show that pulsed laser deposition (PLD) can produce  nanostructures on Ge(001) surfaces which exhibit anisotropic lattice relaxation. Pulsed deposition appears to be advantageous for observation of these effects because the high instantaneous flux promotes the simultaneous nucleation of  small islands that interact elastically at a very early stage and subsequently coarsen via strain-modified ripening. Observation of this process  gives insight into the  effects of anisotropic stress on the structures formed.

The growth and annealing is monitored by real-time Grazing Incidence Diffraction (GID) around the (1 0 0.05)$_s$ surface-sensitive reflection using  synchrotron radiation with wavelength $\lambda$=0.124 nm. Reflections are indexed in a surface unit cell with basis vectors related to the cubic axis as follows: $[1 0 0]_s=(1/2)[1 \bar{1} 0]_{bulk};[0 1 0]_s=(1/2)[1 1 0]_{bulk};[0 0 1]_s=[0 0 1]_{bulk}$. The sample preparation and experimental setup details are given in a supplementary file.\cite{supplementary}

The as-prepared vicinal Ge(001) surface consists of $\sim$ 50 nm wide  terraces with the dimer orientation and stress tensor rotated 90$^\circ$ across each mono-atomic step, as illustrated in Fig. \ref{fig:sub_ML}(a). The large lattice relaxation concentrates on small islands which are free to relax in two dimensions under the constraint of the covalent bonds to the substrate. The x-ray diffraction vector $\vec{Q}$ is oriented along the miscut, as shown.

Fig. \ref{fig:sub_ML}(b) shows the real-time GID intensity evolution near the (1 0 0.05)$_s$ reflection during low temperature growth at 100$^\circ$C.  Selected diffraction profiles are displayed in Fig. \ref{fig:sub_ML}(c). Prior to growth, the diffraction profile shows only  the main diffraction peak, labeled as B (Bulk), as expected for a well-ordered smooth surface.  The sudden drop in intensity of peak B during the deposition burst is related to the formation of a high density of very small islands distributed over the surface.\cite{Vasco2008, Ferguson2009}

During the recovery time after the first deposition burst, a broad diffuse scattering background evolves into two displaced peaks E (Expanded) and C (Contracted) appearing on each side of peak B. Their integrated peak intensities without background subtraction are shown in Fig. \ref{fig:sub_ML}(b) as a function of time. This relaxation behavior is evidence for ripening as has been predicted for PLD.\cite{Vasco2008} In the ripening process, small islands shrink until they disappear, while large islands grow at their expense. Similar effects have  been observed for SrTiO$_3$ homoepitaxy during PLD by specular x-ray scattering, which is sensitive to the island size and correlations.\cite{Ferguson2009}  Since we observe the effect with GID, our experiment is also sensitive to the lattice relaxation of the islands. 

The presence of two distinct diffuse peaks E and C suggests that the coarsened islands exhibit  lattice relaxation relative to the bulk value with both expansion and contraction.   In particular, peak E  is shifted down from the bulk position by  0.033 reciprocal lattice units (rlu) as indicated by curve 2, consistent with lattice expansion of islands atop T$_A$ terraces in the direction parallel to $\vec{Q}$.  The relaxation is reduced slightly in subsequent curves and stabilizes at an average strain of about 2.5$\%$ expansion along the dimer rows.  At 0.4 monolayer (ML) coverage, as shown in curve 4 of Fig. 1(c), the width of peak E along the radial direction ($\delta h$ 0 0)$_s$ is 0.039 rlu. We interpret this width as domain broadening, so that it corresponds to an island size of $L =$10.3 nm along the dimer row direction.  We conclude that the stabilization of both the island size and relaxation is due to the island size approaching a critical size, $L_{c}$, where further ripening is inhibited due to an increase in strain energy. Thus, the largest island that can exhibit nearly complete relaxation is $L_{c} \approx 10$ nm.  

Peak C is weaker, presumably due to the fact that the density of dimer islands on the $T_B$ terraces is less than on $T_A$ terraces, resulting from the anisotropy in surface diffusion and bonding.\cite{Voigtlander2001}  We also note that other effects can also influence the ratio of intensities, such as a small external stress on the surface, which can favor either the (2$\times$1) or (1$\times$2) at the expense of the other.\cite{Men1988} 

\begin{figure}[htbp]
   \centering
   \includegraphics[width=3.2 in]{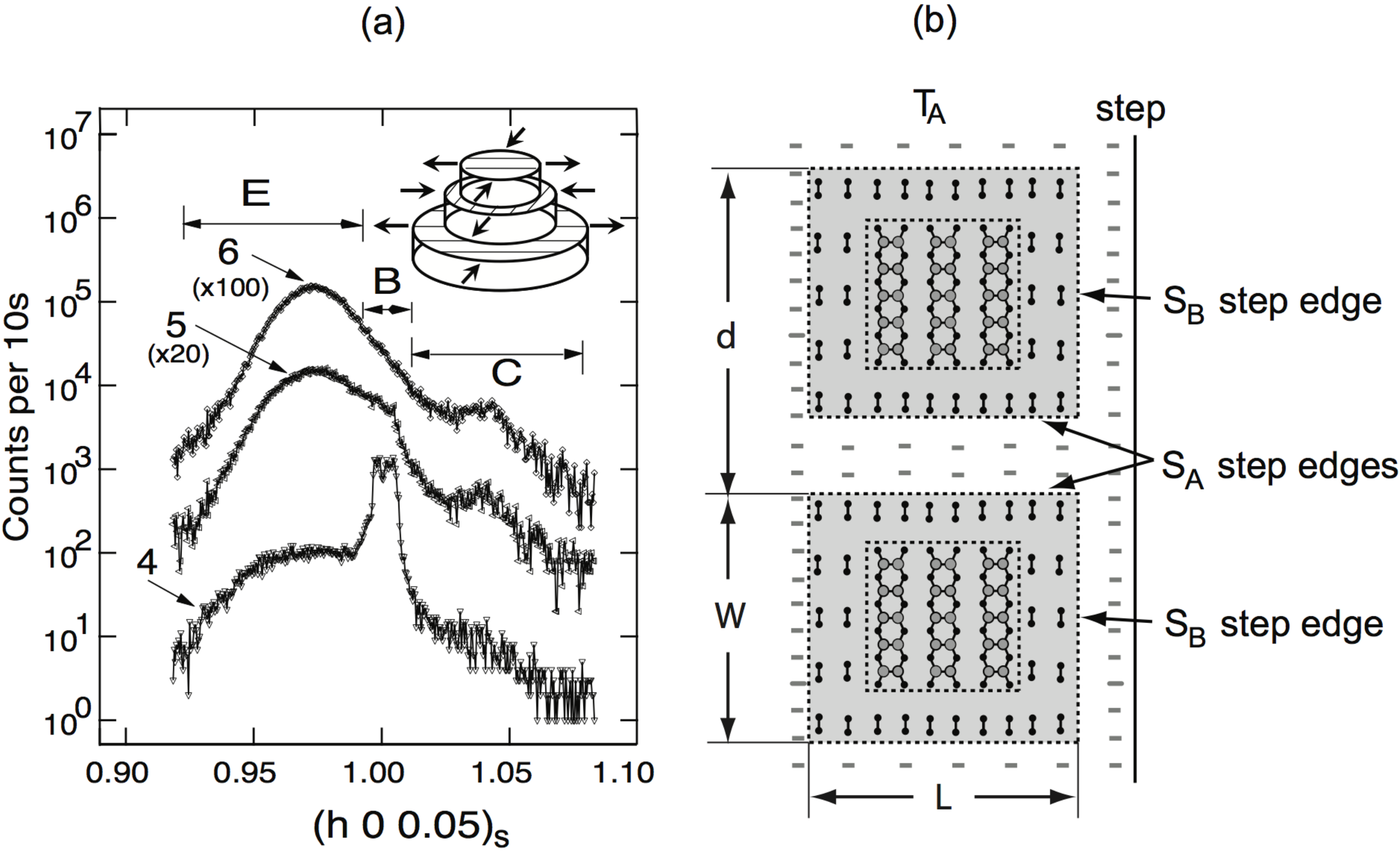}
   \caption{\footnotesize Evolution of the (1 0 0.05)$_s$ reflection in the multilayer growth regime. (a) The data is a continuation of the same deposition run shown in Fig. 1. The three curves correspond to deposited thicknesses of 0.4 (curve 4), 2.0 (curve 5) and 4.0 ML (curve 6), respectively. The curves are shifted vertically for clarity.  Inset: schematic of a multilevel 2D dimer island illustrating the multilevel stress domain model. The wedding-cake-type stacks of (001) terraces are separated by mono-atomic height steps. The dimer rows and the resulting stress anisotropy rotates by 90$^\circ$ at each level, which is illustrated by the lines on each terrace and the arrows along the island periphery at each level. (b) Definitions of length $L$, width $W$, and island spacing $d$, as discussed in the main text.}
   \label{fig:growth}
\end{figure}

It is interesting to consider that all surfaces with anisotropic stress are in principle unstable to formation of elastic-stress domains.\cite{Alerhand1988, Ibach1997} For the case of Ge(001)-(2$\times$1), Middel et al. have found from measurements of the shape of vacancy islands that the stress anisotropy is ($\sigma_{\parallel} - \sigma_{\perp}$) = 8.0 $\pm$ 3.0 eV/nm$^2$.\cite{Middel2002}  Zandvliet et al.  give step free energies at 100$^\circ$C of  $F_{wall}$= 0.048 eV/nm  for $S_A$ step edges and 0.104  eV/nm for $S_B$ step edges. These values allow us to calculate the period $\lambda$ for both orientations of stress domains from

\begin{equation}
\label{eq:myequation}
\lambda = \frac{2\pi a}{\sin(\pi p)} \exp^{(F_{wall}/C + 1)} 
\end{equation} \\

\noindent where $C = (1-\nu)(\sigma_{\parallel} - \sigma_{\perp})^2/2\pi\mu$, $\nu$ is Poisson's ratio, and $\mu$ is Young's modulus.\cite{Alerhand1988, Ibach1997}  Taking the domain fraction $p$ to be $1/2$ and the 1$\times$1 lattice constant to be $a = 0.40$ nm, we have $\lambda_A = 22$ nm and $\lambda_B = 86$ nm.   However, stress domains do not spontaneously form on the initial surface at the predicted length scales.  We suggest  that since there is a nucleation barrier to the formation of small islands the effect of the terrace steps keeps the adatom density on the terraces low enough to retard spontaneous nucleation of stress domains. In this case, stress domains should spontaneously form only if the terrace size can be made significantly larger, as has been observed for Boron doped Si(001).\cite{Ermanoski2011}  Below, we  show from x-ray rocking scans that spatially correlated  islands are formed in PLD, with spacings that closely agree with the values calculated from Eq. (\ref{eq:myequation}), lending credence to the idea that the pronounced strain relaxation plays a role in stabilizing the structures observed.  First, we will continue our discussion of the time-resolved data, now turning our attention to the $> 1$ ML regime.

Fig. \ref{fig:growth} is a continuation of the  deposition run shown in Fig. \ref{fig:sub_ML}, showing additional curves for growth thickness greater than 1.0 ML. Curve 4 is the same as in Fig. \ref{fig:sub_ML}(c) for deposited thickness of 0.4 ML, and curves 5 and 6 are for thickness of 2.0 and 4.0 ML, respectively. The main trend in this growth regime is that peak B intensity continues to decrease and is eventually obscured by the diffuse intensity, while peaks E and C continue to increase. At 4.0 ML coverage (curve 6), diffuse peaks E and C  suggest an average lattice relaxation of about 2.5$\%$ in expansion and 4.0$\%$ in contraction. After the completion of 4.0 ML growth,  $\emph{in-situ}$ x-ray reflectivity shows a root-mean-square roughness of 0.28 nm ($\sim$2 ML), confirming that the growth is multilevel.

This behavior is  contrary to what would be expected if 2D islands grow beyond $L_{c}$, since the lattice spacing on the interior of larger islands would be constrained to the bulk value. Moreover, in many cases, the in-plane lattice parameter is observed to oscillate because small islands become laterally constrained once they coalesce into a continuous layer.\cite{Massies1993, Eymery1994JAP, Hartmann1998}. We do not observe oscillatory relaxation related to 2D island coalescence because multilevel growth takes place, i.e. nucleation of the second monolayer occurs at $\sim$ 0.5 ML and its stress anisotropy, which is rotated by 90$^\circ$ with respect to the layer beneath it, stabilizes the interior of the growing island. Each successive terrace in a given multilevel island is smaller than the terrace beneath it, thus a wedding-cake-type structure is formed, as illustrated in  Fig. \ref{fig:growth}. The dimer direction and corresponding elastic force direction correspondingly rotate across each mono-atomic step gives rise to the alternating contributions to scattering intensity E and C. Thus, we infer the existence of metastable stress domains in multilevel growth. We refer to them as ``metastable" since they are not believed to be lowest energy structures as compared to the two-level (2$\times$1)/(1$\times$2) stress domain envisioned by Alerhand et al.\cite{Alerhand1988} This model most naturally explains the observation that the strain relaxed peaks increase in intensity up to at least 4 ML deposited thickness without broadening significantly. 

Real space images of wedding-cake type multilevel islands have previously been observed in Scanning Tunneling Microscope images  after Ge growth on Ge(001) by Molecular Beam Epitaxy (MBE).\cite{vannostrand1998} Our results are in striking contrast to  the case of Ge MBE where layer-by-layer growth as a result of island coalescence precedes a transition to multilevel growth\cite{supplementary,vannostrand1998}  As a result, for the case of MBE significantly  thicker layers are required to observe the wedding-cake type structures. We explain these differences as follows: At a given growth temperature the nucleation density in MBE  is significantly lower than for the case of PLD because the peak incident particle flux $F$  is orders of magnitude lower, and the spacing between nuclei varies as $(D/F)^{1/6}$, where $D$ is the surface diffusion constant.\cite{Pimpinelli1998} Island sizes may exceed $L_{c}$,  and the monomer density on the surface is subsequently very low, so that nucleation of new small islands is kinetically blocked.  In contrast, for PLD, due to the  high island nucleation density, neighboring islands start to interact with each other through  elastic interaction at an early stage, thus favoring size selection and ordering.  Viewed another way, we can also say that the islands are at the right length scale to break the nucleation barrier for the formation of stress domains.   Moreover, when strain fields of neighboring islands overlap, coalescence can  be inhibited in favor of multilevel growth. This effect has been observed in Monte Carlo simulations.\cite{scholl1998} Alternate models based on standard Ostwald ripening do not predict the formation of regular arrays of equally spaced islands.\cite{scholl1998}   

%\begin{figure}[htbp]
%   \centering
%   \includegraphics[width=3.5 in]{./Fig_3_Surface_Diagram_V5}
%   \caption{\footnotesize Schematic of a minimal four dimer island (DI).  The island itself and the 12 surface atoms beneath the dimers are shown with anisotropic relaxation consistent with our experimental results.  Relaxation of surrounding surface atoms and subsurface atoms down to layer 3 were calculated by minimization of elastic energy using the Keating model.  These atoms are shown in their unrelaxed positions, but with arrowheads indicating the direction of relaxation.  Typical displacements are 2-4 $\%$ of the 1$\times$1 unit cell size (black arrows), and $\approx$0.5 $\%$ (gray arrows).}
%   \label{fig:surfaceschematic}
%\end{figure}

%Fig. \ref{fig:surfaceschematic} illustrates the resulting displacements for the specific case of an isolated  four dimer island (shown as layer DI) atop a 1x1 terminated surface layer (layer 1).

The large lattice relaxation during Ge(001) homoepitaxy illustrated in Figs. \ref{fig:sub_ML} and  \ref{fig:growth} is consistent with a recent observation of 3.6$\%$ contraction in the dimer bond direction near the S$_A$ step edge on Si(001) by non-contact atomic force microscopy at 5 K.\cite{Naitoh2010} It is found that the elastic relaxation extends to about 4.0 nm away from the step. If we consider a 2D island edge instead of a straight step edge, the corresponding critical diameter above which the islands can no longer be fully relaxed would be around $L_{c} =$ 8.0 nm. This predicted island size agrees well with our observation of $\approx$ 10 nm island size at 0.4 ML. 

\begin{figure}[htbp]
   \centering
   \includegraphics[width=3.2 in]{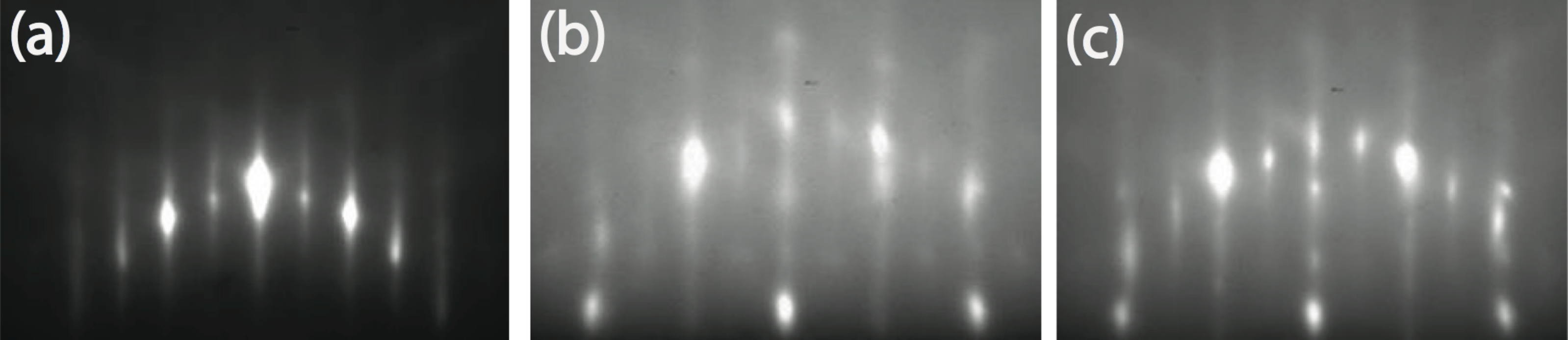}
   \caption{\footnotesize RHEED patterns obtained along the [100]$_s$ azimuth of (a) as-prepared vicinal Ge(001) surface; (b) 4.0 ML thick layer grown at 100$^\circ$C; (c) 5.2 ML thick layer grown at 250$^\circ$C. The electron energy used is 20 keV with incident angle of 2.2$^\circ$, 3.0$^\circ$ and 2.8$^\circ$, respectively.}
   \label{fig:RHEED}
\end{figure}

The contraction resulting in the C peak in our data is associated with the dimer bond formation, and the  expansion in the orthogonal direction resulting in the E peak  is driven by tilting of the dimer, since the lower dimer atom tends to spread out and  push the atoms beneath it laterally.\cite{Zandvliet2003}    We calculated subsurface displacements  by minimizing elastic energy with the Keating model. Displacements calculated for a perfect  2$\times$1 surface structure  show elastic relaxation to at least six layer beneath the surface,  consistent with previous experimental and theoretical results.\cite{pedersen_1989}   Calculations for a minimal four dimer island suggest that the dimer layer and at least two layers beneath the dimer island have large displacements ($> 1\%$), and should contribute to both E and C peaks.   The calculation also shows that large displacements  propagate into the surrounding substrate surface layer. This is consistent with the idea that strain relaxation may influence attatchment/detachment processes at the island edges, as well as introducing a bias to diffusion near the islands. These are possible atomic scale mechanisms to explain the observation that island sizes saturate at $L_{c}$ during ripening.    Both processes have also been suggested to play a role in the ordering of self-organized quantum dots.\cite{pan_2010, meixner_2003}

To link the  film morphology with results found in the literature, Reflection High Energy Electron Diffraction (RHEED) patterns were recorded along the [100]$_s$ azimuth before and after growth. Fig. \ref{fig:RHEED} shows several examples. On the clean surface before deposition, in  (a)  bright spots coexist with half-order reconstruction streaks, indicating a smooth and well-ordered Ge(001)-(2$\times$1) reconstructed surface.\cite{Horn1994} After finishing growth, in RHEED patterns (b) and (c) streaks broaden, half-order streak intensities decrease while diffuse scattering is observed to increase, consistent with the surface roughness deduced from x-ray reflectivity data. In addition, intensity modulations become visible along the integer diffraction streaks. This behavior has previously been correlated with the formation of multilevel 2D islands on the growing surface,\cite{Xue1993, Bratland2003} i.e., the growth mode in which several levels separated by mono-atomic height steps are exposed on the surface.

\begin{figure}[htbp]
   \centering
   \includegraphics[width=3.2 in]{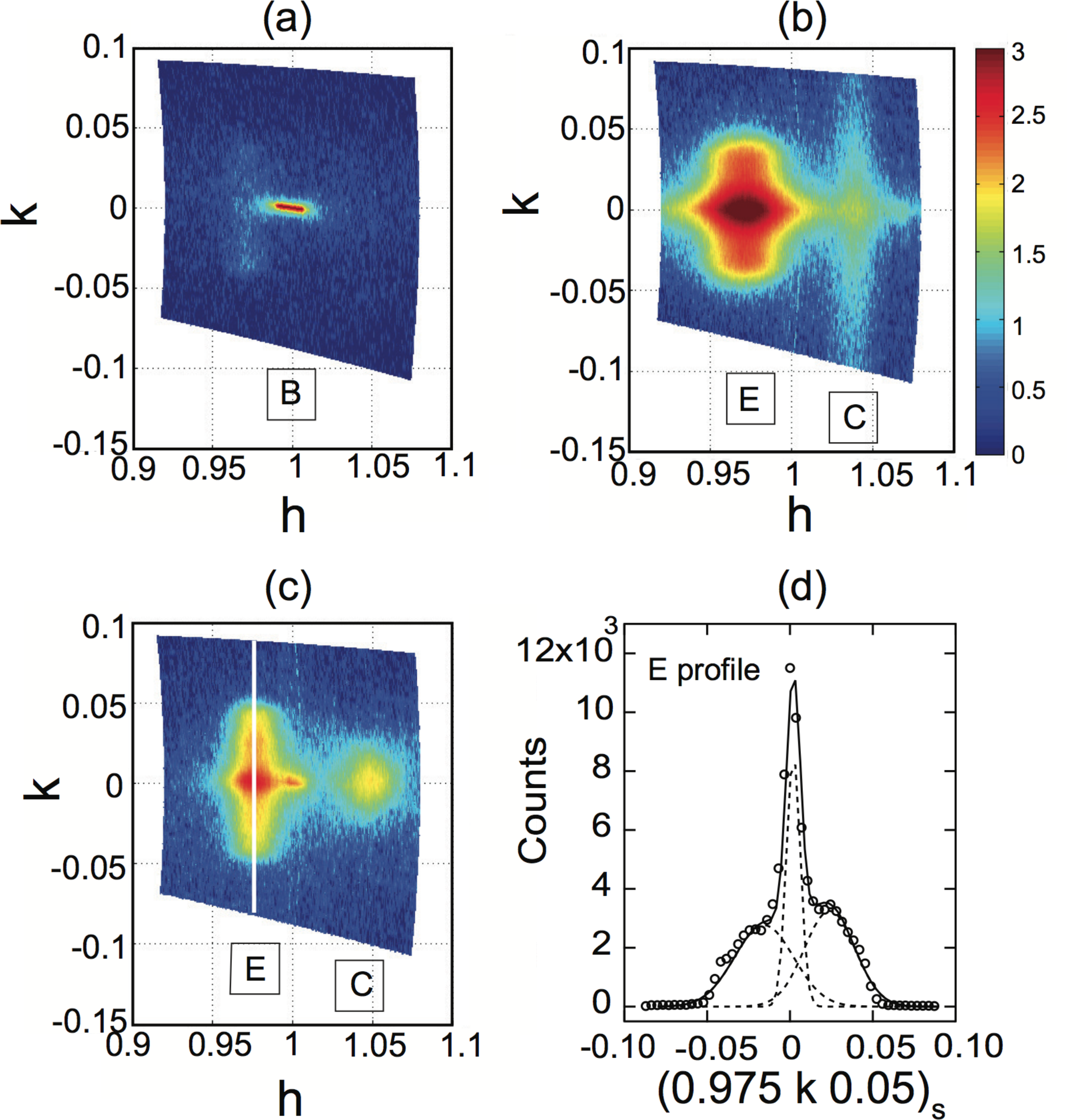}
   \caption{\footnotesize (a)-(b) Reciprocal space maps near (1 0 0.05)$_s$ of the sample deposited at 100$^\circ$C: (a) prior to growth and  (b) after 4.0 ML growth. Plots (c) and (d) show results for a new sample with a deposition temperature of 250$^\circ$C  and subsequent annealing at 650$^\circ$C for 400 s and cooling down to room temperature. Rocking scans have been converted from angles to reciprocal lattice units. The line profile in (d) is along the transverse direction (0 $\delta k$ 0)$_s$ for peak E, as indicated by the vertical white line in (c). The dotted lines are fitted Gaussian profiles.}
   \label{fig:sample3_anneal}
\end{figure}

In order to obtain additional information about the multilevel island structure, we performed rocking scans on the sample about its surface normal. A linear position sensitive detector was used so that 2D reciprocal space maps can be rapidly obtained with a single scan. We first describe how lattice relaxation, island size broadening and island correlation affect the GID data: (i) In reciprocal space near (100)$_s$, radial ($\delta h$ 0 0)$_s$ intensity profiles are sensitive to the strain relaxation parallel to the surface.  The peak position indicates the average relaxation, while the peak width can be limited by the distribution of lattice constants or by domain size broadening. In the latter case, the width is approximately $a/L$, where the surface unit cell constant is $a$ = 0.400 nm and $L$ is the island size along $\vec{Q}$ as shown in Fig. \ref{fig:growth}.  (ii) Transversely, (0 $\delta k$ 0)$_s$ diffuse scattering gives information about the extent of correlations laterally across the surface. In cases where islands are randomly distributed on the surface, the diffuse scattering will be centered at $k$ = 0 and its width will be  approximately $a/W$,  where $W$ is the average island size. (iii) In cases where island positions are spatially correlated,  satellite peaks occur near $\Delta k = \pm a/d$, where $d$ is the average island-island spacing.\cite{Guinier1995} 

Fig. \ref{fig:sample3_anneal} (a)-(b) shows detailed reciprocal space maps at the (1 0 0.05)$_s$ position of the sample (a) prior to growth, (b) after  4.0 ML growth at 100$^\circ$C. In Fig. \ref{fig:sample3_anneal}(a), a high intensity peak B indicates a well-ordered starting surface.  In Fig. \ref{fig:sample3_anneal}(b), peak B is no longer visible, due to the surface roughening. Two satellite peaks are observed vertically on each side of peak E, near $\Delta k=-0.018$ and $+0.021$ rlu. This indicates a strong spatial correlation of multilevel 2D islands along the dimer bond direction with a uniform island spacing of $d \approx$ 20 nm.  The effect is anisotropic because no correlation is observed in the orthogonal direction. This value of  $d$ is in close agreement with the length scale $\lambda_A$ calculated for stress domains, and is reasonable if we assume that the length scale was set during the submonolayer stage of growth, given that the islands are separated by $S_A$ steps, as shown in Fig. \ref{fig:growth}(b). We also find that $L \approx$ 16.9 nm.  To complete our description of  Fig. \ref{fig:sample3_anneal}(b), we note that the widths of peak C  indicate an average island size of $L \approx$ 15.1 nm along the dimer direction, and  $W \approx$ 12.3 nm in the transverse  direction.  Since peak C only increases rapidly for depositions $ > 1$ ML, we interpret these widths as being predominantly relevant to domains within multilevel islands atop $T_A$ terraces. 

Additional information about surface kinetics can be obtained by annealing the metastable multilevel structures discussed so far.    Fig. \ref{fig:sample3_anneal} (c)  shows an additional reciprocal space map on another sample deposited at 250$^\circ$C with a total thickness of 5.2 ML, followed by annealing at 650$^\circ$C. It is noted that  the new sample shows almost the same peak E and C positions and widths as the one deposited at 100$^\circ$C. This shows that the island size again saturates near $L_{c}\approx 10$ nm for submoloayer deposition, and that it is a true saturation rather than simply a result of the kinetics slowing down as the island size increases because the coarsening process is much faster at 250$^\circ$C. Upon annealing peak positions stay unchanged. However, the profile of peak C becomes more compact along the $k$ direction, while peak E becomes more compact along the radial $h$ direction. Specifically,  peak E changes  width along $h$, corresponding to a change in the island size along the dimer row direction from $L \approx$ 12.5 to 20.0 nm.  Peak C also becomes more compact in $k$, suggesting that $W$ changes from $<10$ to 16.7 nm. The satellite peaks for diffuse scattering E shown in Fig. \ref{fig:sample3_anneal}(d) after annealing are located at $\Delta k = \pm0.019$ rlu, showing that the island spacing along the dimer bond direction has changed slightly from $D \approx$ 18.8 nm to 21.0 nm, indicating that minimal ripening has occurred and hence the majority  of the mass transport is related to transport within the multilevel islands. 

We have annealed stepwise to higher temperatures until the metastable islands relax away and the surface returns to a nearly flat state at 800$^\circ$C. In a separate set of experiments, we also confirmed that the surface phase transition occurs at $\approx 780^\circ$C.  The transition is characterized by reversible broadening of the (3/2 0 0.05) reflection, which disappears entirely above the transition temperature, and loss of the ordered terrace structure, as confirmed by monitoring the peak splitting in $k$ scans at  (0 1 0.05). Our observations are consistent with the suggestion of Zandvliet et al.,\cite{Zandvliet2000} that the transition is due to a roughening transition where the step edge free energy $F_{wall}$ is reduced to zero rather than by breaking of dimer bonds. In this case, Eq. \ref{eq:myequation} predicts a length scale of $\lambda \approx 7$ nm with no anisotropy, indicating that small 3-4 nm islands may still be present on the surface during annealing at 800$^\circ$C, but these are not directly detected in our measurements.

In conclusion, we have observed two main effects: (i) anisotropic relaxation in submonolayer deposition during pulsed growth with saturation of island sizes at $\sim$10 nm, and (ii) persistence of anisotropic relaxation during multilevel growth and subsequent annealing to moderate temperatures. The results show that  high density nucleation followed by evolution through a ripening process leads to a narrow distribution of size-selected islands, which can also be viewed as breaking the kinetic barrier to the formation of stress domains.   

The authors acknowledge Christie Nelson and Steve Lamarra for experimental assistance with the work done at the NSLS X21 beamline. Research supported by the U.S. Department of Energy, Office of Basic Energy Sciences, Division of Materials Sciences and Engineering under Award No. DE-FG02-07ER46380. Use of the National Synchrotron Light Source was supported by the U.S. Department of Energy. Development of the capability for $\emph{in-situ}$ x-ray analysis of PLD  was supported by the National Science Foundation under DMR-0216704 and DMR-0348354.

% tables should appear as floats within the text
% The ruledtabular enviroment adds doubled rules to table and sets a
% reasonable default table settings.
% Use the table* environment to get a full-width table in two-column
% Add \usepackage{longtable} and the longtable (or longtable*}
% environment for nicely formatted long tables. Or use the the [H]
% placement option to break a long table (with less control than
% \begin{table}%[H] add [H] placement to break table across pages
% \begin{ruledtabular}
% Lines of table here ending with \\
% \end{tabular}
% \end{ruledtabular}
% \end{table}

% Surround table environment with turnpage environment for landscape
% table
% \begin{table}
% \begin{ruledtabular}
% \end{tabular}
% \end{ruledtabular}
% \end{table}

%\appendix
%\section{}

%\begin{acknowledgments}

%\end{acknowledgments}

\bibliography{Ge_PLD_V28}

\end{document}